\documentclass[twocolumn]{revtex4}
\usepackage{amssymb}
\usepackage{amsmath}             % Using a mathematical package
\usepackage[dvips]{graphicx,color}

\begin{document}

\title{Phase-space approach to the study of decoherence in quantum walks}
\author{Cecilia C. L\'opez$^1$ and Juan Pablo Paz$^{1,2}$}
\affiliation{(1): Departamento de F\'\i sica ``J.J. Giambiagi'', FCEyN UBA,
Pabell\'on 1, Ciudad Universitaria, 1428 Buenos Aires, Argentina}
\affiliation{(2): Theoretical Division,  MSB210, Los Alamos National Laboratory,
Los Alamos, NM 87545}
\begin{abstract}
We analyze the quantum walk on a cycle using discrete Wigner functions as a way to represent
the states and the evolution of the walker. The method provides some insight on the nature
of the interference effects that make quantum and classical walks different. We also study the
behavior of the system when the quantum coin carried by the walker
interacts with an environment. We show that for this system quantum coherence is robust
for initially delocalized states of the walker. The use of phase-space representation enables us
to develop an intuitive description of the nature of the decoherence process in this system.
\end{abstract}
\date{September 4, 2003}


\maketitle

\section{Introduction}

Quantum walks \cite{Kempe} have been proposed as potentially useful components
of quantum algorithms \cite{Aharonov}. In recent years these systems have been
studied in detail and some progress has been made in developing new quantum algorithms
using either continuous \cite{ChildsExp} or discrete \cite{ShenviExp} versions of
quantum walks. The key to the potential success of quantum walks seems to rely on the
ability of the quantum walker to efficiently spread over a graph (a network of sites) in
a way that is much faster than any algorithm based on classical coin tosses.

Quantum interference plays an important role in quantum walks being the crucial ingredient
enabling a faster than classical spread. For this reason, some effort was made
in recent years in trying to understand the implications of the process of decoherence
for quantum walks \cite{KendonDec,BrunOct,KendonJan}. Decoherence, an essential ingredient
to understand the quantum--classical transition \cite{DecoRef}, could turn the quantum
walk into an algorithm as inefficient as its classical counterpart. The models studied in this
context can be divided in two classes depending on how the coupling with an external
environment
is introduced. In fact, a quantum walk consists of a quantum particle that can occupy a
discrete set of points on a lattice. In the discrete version, the walker carries
a quantum coin, which in the simplest
case can be taken as a spin-$1/2$ degree of freedom. The algorithm proceeds so that the walker moves
in one of two possible directions depending on the state of the spin (for more complex regular
arrays, a higher spin is required).
So, in this context it is natural to consider some decoherence models where the spin is
coupled to the environment and others where the position of the walker is directly coupled
to external degrees of freedom. The specific system in which the algorithm is implemented
in practice will dictate which of these two scenarios is more relevant. Several experimental
proposals to implement discrete quantum walks in systems such as ion
traps \cite{Travaglione}, cavity QED \cite{cavityQED}, and optical lattices \cite{Dur}
have been analyzed (see also Ref. \cite{Du} for a recent NMR implementation of a
continuous quantum walk).

The main effect of decoherence on quantum walks is rather intuitive:
as the interaction with the environment washes out quantum interference effects,
it restores some aspects of the classical behavior. For example, it has been shown
that the spread of the decohered walker becomes diffusion dominated proceeding slower
than in the pure quantum case. This result was obtained both for models with decoherence
in the coin and in the position of the walker \cite{KendonDec,BrunOct,KendonJan}.
However, it is known that classical correspondence in these systems has some surprising
features. For example, for models with some decoherence in the quantum coin the asymptotic
dispersion of the walker grows diffusively but with a rate that does not coincide with
the classical one \cite{BrunOct}. Also, a small amount of decoherence seems to be useful
to achieve a quantum walk with a significant speedup \cite{KendonDec,KendonJan}.

In this work we will revisit the quantum walk on a cycle (and on a line) considering
models where the quantum coin interacts with an environment. The aim of our work is
twofold. First we will use phase-space distributions (i.e., discrete Wigner
functions) to represent the quantum state of the walker. The use of such distributions
in the context of quantum computation has been proposed in Ref. \cite{MPS02}, where some general
features about the behavior of quantum algorithms in phase space were noticed. A phase-space
representation is natural in the case of quantum walks, where both position and
momentum play a natural role. Our second goal is to study the true nature of the
transition from quantum to classical in this kind of model. We will show that models where
the environment is coupled to the coin are not able to induce a complete transition to
classicality. This is a consequence of the fact that the preferred observable selected
by the environment is the momentum of the walker. This observable, which is the generator of
discrete translations in position, plays the role of the ``pointer observable'' of the system
\cite{DecoRef,Zurek81}. Therefore, as we will see, the interaction with the environment
being very efficient in suppressing interference between pointer states preserves the
quantum interference between superpositions of eigenstates of the conjugate observable to
momentum (i.e., position). Again, the use of phase-space representation of quantum
states will be helpful in developing an intuitive picture of the effect of decoherence
in this context. The paper is organized as follows: In Sec. II we review some basic aspects
of the quantum walk on the cycle. We also introduce there the phase-space representation of
quantum states for the quantum walk and discuss some of the main properties of the discrete
Wigner functions for this system. In Sec. III we introduce a simple decoherence model
and show the main consequences on the quantum walk algorithm. In Sec. IV we present
a summary and our conclusions.

\section{Quantum walks and their phase-space representation}

\subsection{Quantum walks on the cycle}

The quantum walks on an infinite line or in a cycle with $N$ sites are simple enough
systems to be exactly solvable. For the infinite line the exact
solution was presented in Ref. \cite{Nayak}. The case of the cycle was first solved in
Ref. \cite{Aharonov}. However, the exact expressions are involved enough to require numerical
evaluation to study their main features. Here we will review the main properties of this
system presenting them in a way which  prepares the ground to use phase-space representation
for quantum states (we will focus on the case of a cycle, the results for the line can be
recovered from ours with $t\le N$).

For a quantum walk in a cycle of $N$ sites, the Hilbert space
is  $\mathcal{H}=\mathcal{H}_N \otimes \mathcal{H}_2$, where  $\mathcal{H}_N$ is the space
of states of the walker (an $N$-dimensional Hilbert space) and  $\mathcal{H}_2$ is the
two--dimensional Hilbert space of the quantum coin (a spin $1/2$). The algorithm is defined
by a unitary evolution operator which is the iteration of the following map:
\begin{equation}
\mathcal{U}_{walk} = U^{\sigma_z}\  H.
\label{onestep}
\end{equation}
Here  $H$ is the Hadamard operator acting on the Hilbert space of the quantum coin
[$H=(\sigma_x+\sigma_z)/\sqrt{2}$, $\sigma_i$ being the usual $2\times 2$ Pauli matrices].
The operator $U$ is the cyclic translation operator for the walker, which in the
position basis is defined as $U|n\rangle = |n+1\rangle$ (mod $N$). It is worth
stressing that the operator $U^{\sigma_z}$ is nothing but a spin--controlled translation
acting as
$U^{\sigma_z}|n\rangle\otimes|{0\atop 1}\rangle = |n\pm1\rangle\otimes|{0\atop 1}\rangle$.
So, the map $\mathcal{U}_{walk}$ consists of a spin--controlled translation preceded by a
coin toss, which is implemented by the Hadamard operation (the use of the Hadamard operator
in this context is not essential and can be replaced by almost any unitary operator on the
coin \cite{Kempe}).

The notion of phase-space is natural in the context of this kind of quantum walk. In fact,
the position eigenstates $\{|n\rangle,\ n=0,\ \ldots,\ N-1\}$ form a basis of the walkers'
Hilbert space $\mathcal{H}_N$. The conjugate basis is the so--called momentum
basis $\{|k\rangle,\ k=0,\ \ldots,\ N-1\}$. Position and momentum bases are related
by the discrete Fourier transform, i.e.,
\begin{equation}
\langle n|k\rangle = {1\over\sqrt{N}} \exp(2\pi i nk/N).\label{dft}
\end{equation}
The cyclic translation operator $U$ that plays a central role in the quantum walk
is diagonal in momentum basis since
$U|k\rangle=\exp(-i2\pi k/N)|k\rangle$. This simply indicates that momentum is nothing
but the generator of finite translations. As a consequence of this, the total unitary
operator defining the quantum walk algorithm is also diagonal in such basis.
This fact, which was noticed before by several authors, enables a simple exact solution of
the quantum walk dynamics. Indeed, we can write the initial state of the system using the
momentum basis of the walker as (below $\rho$ denotes the total density matrix of the
system formed by the walker and the coin)
\begin{equation}
\rho(0)=\sum_{k,k'=0}^{N-1} c_{k,k'} |k\rangle\langle k'| \otimes \rho_2(0),
\label{roinic}
\end{equation}
where $\rho_2(0)$ is the initial state of the quantum coin (we assume that the initial
state of the combined system is a product, but this assumption can be relaxed).
After $t$ iterations of the quantum map the reduced density matrix of the walker (denoted
as $\rho_w$) is
\begin{eqnarray}
\label{ro1planteo}
\rho_{w}(t) = \sum_{k,k'=0}^{N-1} c_{k,k'} f(k,k',t)\ |k\rangle\langle k'|, \\
\label{fkkplanteo}
f(k,k',t)={\rm Tr}_2[M_k^t \rho_2(0) {M_{k'}^\dagger}^t],
\end{eqnarray}
where the operator $M_k$ is defined as
\begin{equation}
M_k = \exp(-i2 \pi \ k \sigma_z/N) \ H.
\label{Mofk}
\end{equation}

All the temporal dependence is contained in the function $f(k,k',t)$
which can be exactly computed in a straightforward way:
One should first expand the initial state $\rho_2(0)$ in a (nonorthogonal) basis
of operators of the form $|\phi^l_k\rangle\langle \phi^m_{k'}|$ ($l,m=1,2$), where
$|\phi_k^l\rangle$ are the eigenstates
of the operator $M_k$ (i.e., $M_k|\phi_k^l\rangle=\lambda_{k,l}|\phi_k^l\rangle$).
The explicit expressions for the eigenstates $|\phi_k^l\rangle$ and the eigenvalues $\lambda_{k,l}$
will not be given here since they can be found in the literature (see, for example,
\cite{Nayak}). After doing this the evolution of the quantum state is fully determined by the
equation
\begin{equation}
f(k,k',t)=\sum_{l,m=1,2}(\lambda_{k,l}\lambda^{*}_{k',m})^t\
\langle\phi^l_k|\rho_2(0)|\phi^m_{k'}\rangle\
\langle\phi^m_{k'}|\phi^l_k\rangle.
\end{equation}
Below we will describe the properties of this solution using
a phase-space representation for the quantum state of the system.

\subsection{Phase-space representation}

Wigner functions \cite{Wigner} are a powerful tool to represent the state and the
evolution of a quantum system. For systems with a finite-dimensional
Hilbert space, the discrete version of Wigner functions was introduced
using different methods (see Ref. \cite{Wigner-Discrete}). We will follow the approach
and notation used in Ref. \cite{MPS02}, where these phase-space distributions were applied
to study properties of quantum algorithms. For completeness, we will give here the
necessary definitions and outline some of the most remarkable properties of the
discrete Wigner functions.

For a system with an $N$-dimensional Hilbert space the discrete Wigner function can be
defined as the components of the density matrix in a basis of operators defined as
\begin{equation}
A(q,p)=U^qR V^{-p}\exp(i\pi pq/N).
\label{aqp}
\end{equation}
These are the so--called phase-space point operators. They are defined in terms of
the cyclic shift $U$ (which in the position basis acts as $U|n\rangle=|n+1\rangle$),
the reflection operator $R$ (which in the position basis acts as $R|n\rangle=|-n\rangle$),
and the momentum shift $V$ (which generates cyclic displacements in the momentum basis,
i.e., $V|k\rangle=|k+1\rangle$).
Phase-space operators are unitary, Hermitian and form a complete orthogonal
basis of the space of operators (they are orthogonal in the Hilbert--Schmidt inner
product since they satisfy that ${\rm Tr}[A(q,p)A(q',p')]=N\delta_{q,q'}\delta_{p,p'}$).
Expanding the quantum state in the $A(q,p)$ basis as
\begin{equation}
\rho=\sum_{q,p=0}^{N-1}W(q,p) A(q,p),
\label{state}
\end{equation}
the coefficients $W(q,p)$ are the discrete Wigner functions of the quantum state, which
are obtained as
\begin{equation}
W(q,p)={1\over N}{\rm Tr}[\rho A(q,p)].
\label{wigner}
\end{equation}
This function has three remarkable properties that almost give it the status of a
probability distribution. The first two properties are evident: Wigner functions are
real numbers (a consequence of the Hermiticity of the phase-space operators) and
they provide a complete description of the quantum state (a consequence of the completeness
of the basis of such operators). The third property is less obvious: marginal probability
distributions can be obtained by adding values of the Wigner function along lines in
phase-space. For this to be possible, it turns out that the phase-space has to be
defined as a grid of $2N\times 2N$ points where $W(q,p)$ is given at each point precisely
by Eq. (\ref{wigner}). Thus, adding the values of the Wigner function over all points
satisfying the condition $aq-bp=c$ one gets the probability to detect an eigenstate
of the operator $D(b,a)=U^bV^a\exp(i\pi ab/N)$ with eigenvalue $\exp(i\pi c/2N)$ (the sum is
equal to zero if such eigenvalue does not exist). In particular, adding the Wigner function
along vertical lines $q=c$ one obtains the probability to detect eigenstates of the
operator $D(0,1)=V$, with eigenvalues given by $\exp(i\pi c/2N)$. These numbers are
equal to zero if $c$ is odd and they are equal to the probability for measuring the
position eigenstate $|c/2\rangle$ when $c$ is even. Complementary, adding values
of the Wigner function along horizontal lines enables us to compute the probability
to detect a momentum eigenstate.

\begin{figure}[hpt]
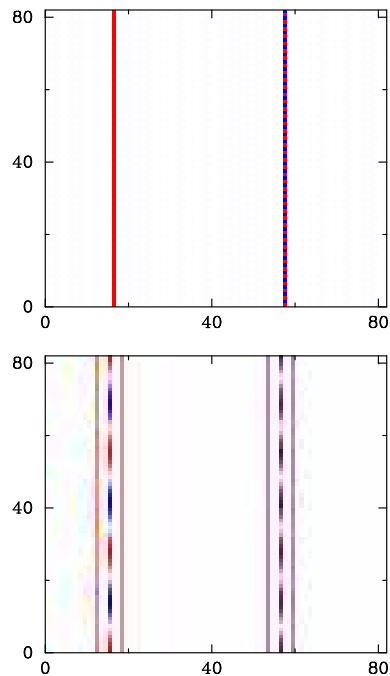

\begin{center}
\includegraphics[angle=270,width=0.28\textwidth]{fig1a.eps}
\\
\vspace{0.3cm}
\includegraphics[angle=270,width=0.28\textwidth]{fig1b.eps}
\end{center}
\caption{
Wigner functions $W(q,p)$ for a localized state (up) and for a delocalized superposition
state (down). The dimension of the Hilbert space is $N=41$. The horizontal (vertical) axis
corresponds to position (momentum). Color code is such that red (blue) regions correspond to
positive (negative) values while white corresponds to zero ($n_0=8$, $n_1=6$, $n_2=9$).}
\label{wigner-basicas}
\vspace{0.2cm}
\hrule
\end{figure}

A final remark about properties of the discrete Wigner function is in order. Figure \ref{wigner-basicas}
shows the Wigner function of a position eigenstate $|n_0\rangle$ and of a superposition of
two position eigenstates, such as $(|n_1\rangle + |n_2\rangle)/\sqrt{2}$.
As we see, in the first case the function is positive on a
vertical line located at $q=2n_0$ and is oscillatory on a vertical line located at
$q=2n_0+N$. The interpretation of these oscillations is clear. It is well known that
Wigner functions display oscillatory regions whenever there is
interference between two pieces of a wave packet. In this case, the cyclic boundary
conditions we are imposing (that originate from the fact that $U$ and $V$ are cyclic
shift operators) generate a mirror image for every phase-space point. Thus, the oscillating
strip can be interpreted as the interference between the positive strip and its mirror
image. For the case of a quantum state which is a superposition of two position
eigenstates, we observe two positive vertical lines with the usual interference fringes
in between them. All these vertical lines have their corresponding oscillatory counterparts
originated from the boundary conditions which are located at a distance $N$. In what follows
we will show Wigner functions for typical states of a quantum walker.

\begin{figure}[htp]
\begin{center}
\includegraphics[angle=270,width=0.3\textwidth]{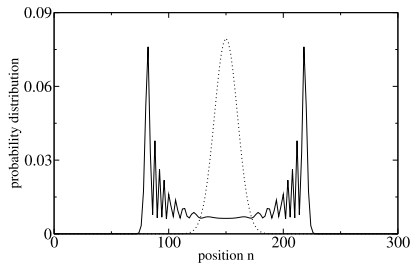}
\end{center}
\caption{Probability distribution in position after $t=100$ iterations for an initially
localized walker with unbiased spin. $N=301$, $n_0=150$. We only plot
the function for sites such that $n+t+n_0$ adds to an even number (solid),
and also include the classical distribution (dotted).}
\label{probab-nodeco}
\vspace{0.2cm}
\hrule
\end{figure}

\begin{figure*}[htp]
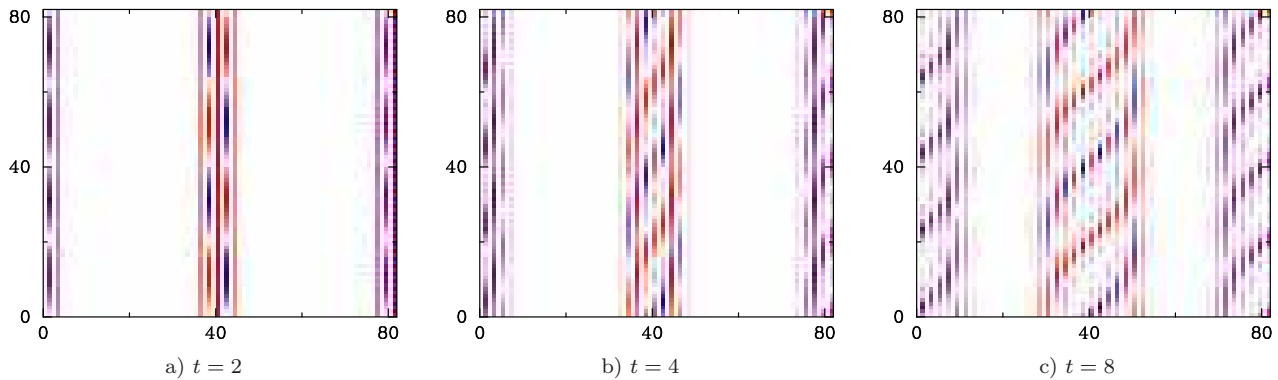

\begin{center}
\begin{minipage}[b]{0.29\textwidth}
\centering
\includegraphics[angle=270,width=1.0\textwidth]{fig3a.eps}\\
\vspace{0.2cm}
\footnotesize
a) $t=2$
\end{minipage}%
\hspace{0.5cm}
\begin{minipage}[b]{0.29\textwidth}
\centering
\includegraphics[angle=270,width=1.0\textwidth]{fig3b.eps}\\
\vspace{0.2cm}
\footnotesize
b) $t=4$
\end{minipage}%
\hspace{0.5cm}
\begin{minipage}[b]{0.29\textwidth}
\centering
\includegraphics[angle=270,width=1.0\textwidth]{fig3c.eps}\\
\vspace{0.2cm}
\footnotesize
c) $t=8$
\end{minipage}%
\end{center}
\begin{flushleft}
%\vspace{0.2cm}
\footnotesize
\caption{Discrete Wigner function $W(q,p)$ at different times. The initial state of the
walker is localized at $n_0=20$ (the dimension of the Hilbert space is $N=41$). The
quantum coin is in an unbiased initial state. Horizontal (vertical) axis corresponds
to position (momentum).}
\label{wigner-local-nodeco}
\vspace{0.2cm}
\hrule
\end{flushleft}
\end{figure*}

For an initial state where the walker starts at a given position and the spin is
initially unbiased [$|\psi(0)\rangle_2 = (|0\rangle +i|1\rangle)/ \sqrt{2}$], the behavior
of the quantum walker starts to deviate from its classical counterpart at early times
(in this paper we will only consider unbiased initial states for the quantum coin). The
phase-space representation of the state is shown in Fig. \ref{wigner-local-nodeco}
and makes evident that
a peculiar pattern of quantum interference fringes develops between the different pieces
of the wave packet. The consequence of these interference effect is evident also when one
looks at the probability distribution for different positions. This distribution is shown
in Fig. \ref{probab-nodeco} and has been previously studied in the literature (see
\cite{Dur,Nayak,BrunOct,KendonDec}). Like its classical counterpart, at a given time $t$ the state
initially located at $n_0$ has support only on states $n$ satisfying that $n+t+n_0$
adds up to an even number. However, in general the quantum distribution differs from the
classical one, exhibiting peaks located at $n=\pm t/\sqrt{2}$ and a plateau
of height $1/\sqrt{2}t$ around $n_0$.
After some time the Wigner function of the quantum  walker develops a shape that
resembles a thread, as it is clear in the pictures. For this reason we will call this
a thread state. \\

\begin{figure*}[hptb]
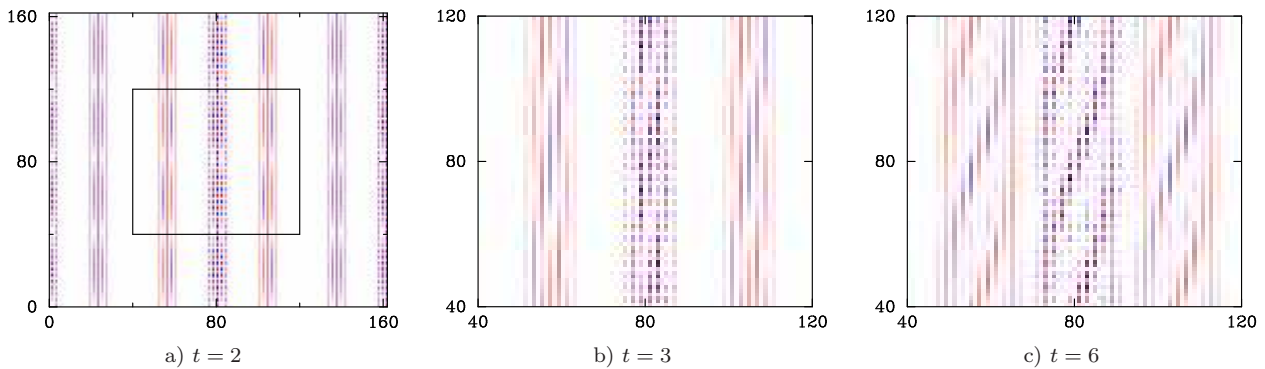

\begin{center}
\begin{minipage}[t]{0.29\textwidth}
\centering
\includegraphics[angle=270,width=1.0\textwidth]{fig4a.eps}\\
\vspace{0.2cm}
\footnotesize
a) $t=2$
\end{minipage}%
\hspace{0.4cm}
\begin{minipage}[t]{0.29\textwidth}
\centering
\includegraphics[angle=270,width=1.0\textwidth]{fig4b.eps}\\
\vspace{0.2cm}
\footnotesize
b) $t=3$
\end{minipage}%
\hspace{0.4cm}
\begin{minipage}[t]{0.29\textwidth}
\centering
\includegraphics[angle=270,width=1.0\textwidth]{fig4c.eps}\\
\vspace{0.2cm}
\footnotesize
c) $t=6$
\end{minipage}%
\footnotesize
\caption{Discrete Wigner function $W(q,p)$ at different times, for an initial state of
the walker which is a superposition of two position eigenstates ($N=81,\ n_1=28,\ n_2=52$)
and an unbiased initial quantum coin. In (a) we plot the complete Wigner function (horizontal
axis and vertical axis correspond to position and momentum, respectively).  In (b) and (c) we
only plot the Wigner function in the phase-space region defined by the black rectangle
shown in (a). In this way the small scale oscillations of the Wigner function
can be observed in detail.}
\label{wigner-delocal-nodeco}
\vspace{0.2cm}
\hrule
\end{center}
\end{figure*}

It is also interesting to analyze the evolution of the quantum walk for delocalized
initial states. In particular, we will consider an initial state that is a coherent
superposition of two position eigenstates (whose Wigner function was already displayed
in Fig. \ref{wigner-basicas}). We find a Wigner function that develops into
a sum of two threads with a region in between where interference fringes are evident.
This is displayed in Fig. \ref{wigner-delocal-nodeco}. Some properties of the quantum walk
for this kind of delocalized initial states were analyzed in Ref. \cite{Grudka} where it was
noticed that the asymptotic probability distribution can be rather different from the
one obtained from a localized initial state. Below, we will show that the process of
decoherence affects localized and delocalized initial states in a rather different way.

\section{Decoherence and the transition from quantum to classical}

\subsection{The decoherence model}

We will consider a quantum walk where the quantum coin couples to an
environment. To describe such coupling we will use a model which was introduced
and studied in detail in Ref. \cite{Teklemariam}. In that paper it was shown that
one can mimic the coupling to an external environment by a sequence of random
rotations applied to the quantum coin, which have the effect of scrambling
the spin polarization. More precisely, these kicks will be generated by
the evolution operator
\begin{equation}
K_j=\exp(-i\epsilon_j \ \hat{n} \cdot \vec{\sigma}),
\label{definicionKj}
\end{equation}
where the angles $\epsilon_j$ take random values and $\hat{n}$ is a fixed vector
specifying the rotation axis. The virtue of this model is not only its simplicity but
also the fact that can be experimentally implemented in a controllable manner
using, for example, NMR techniques.

After the application of one step of the quantum walk algorithm and one kick the
evolution of the total system is
\begin{equation}
\rho(t+1) = K_j \ U^{\sigma_z} H \ \rho(t) \ H U^{-\sigma_z}\ K_j^\dagger,
\label{onestep_deco}
\end{equation}
To obtain a closed expression for the reduced density matrix of the walker for
an ensemble of realizations of the random variables $\epsilon_j$ we follow the
method proposed in Ref. \cite{Teklemariam} (see Ref. \cite{BrunOct} for a similar approach):
Assuming that these angles are randomly distributed in an interval ($-\alpha$,$+\alpha$),
this density matrix is
\begin{eqnarray}
\nonumber
\rho_{w}(t) &=&{\rm Tr}_2 [ \int_{-\alpha}^{\alpha} {d\epsilon_t \over 2\alpha}
\ldots \int_{-\alpha}^\alpha {d\epsilon_1 \over 2\alpha}
 K_t U^{\sigma_z} H \ldots K_1 U^{\sigma_z} H \\
& & \rho_w(0)\otimes\rho_2(0) \ H U^{-\sigma_z} K_1^\dagger \ldots H U^{-\sigma_z}
K_t^\dagger ]\
\label{rodet}
\end{eqnarray}
Expanding the initial state in the momentum basis as before enables us to simplify
this expression. In fact, after doing this one can integrate over the random variables
to find
\begin{eqnarray}
\label{ro1planteodeco}
\rho_{w}(t) &=& \sum_{k,k'=0}^{N-1} c_{k,k'} \tilde f_n(k,k',t)
|k\rangle\langle k'|,
\label{fkkplanteodeco}\\
\tilde f_n(k,k',t)&=& {\rm Tr}_2[O_n^t (\rho_2(0))].
\label{fkkp}
\end{eqnarray}
Here $O_n$ is a superoperator acting on the spin state (depending on
direction $\hat{n}$ of the kicks) as
\begin{eqnarray}
O_n(\rho_2)&=&{(1+\gamma)\over 2}M_k\rho_2 M_{k'}^\dagger \nonumber\\
&+&{(1-\gamma)\over 2}\sigma_nM_k\rho_2 M_{k'}^\dagger \sigma_n, \label{superop}
\end{eqnarray}
where $\gamma=\sin(2\alpha)/(2\alpha)$ is a parameter
related to the strength of the noise (notice that $\gamma=1$ corresponds to unitary
evolution, i.e., to $\alpha=0$). One can find a simple matrix representation for the
superoperator $O_n$ for different choices of the rotation axis by writing $\rho_2(0)$ in the
basis formed by the identity and the Pauli operators. In the Appendix we show the explicit
form of this matrix representation, which is helpful in finding exact and numerical solutions
to the problem. In what follows we will show results for the case $\hat n=\hat y$ (the other
cases are qualitatively similar).
To find the state of the walker at arbitrary times we simply need to find eigenstates and
eigenvalues of the superoperator $O_n$. This can always be done numerically and also
analytically in the interesting case of $\gamma=0$, which can be denoted as ``total decoherence''.
In such case, the exact solution turns out to be
\begin{eqnarray}
\tilde f_y(k,k',t) &=& \cos^t(2\pi(k-k')/N)\nonumber\\
&\times&\left (1 -i\ p_x {\sin(2\pi (k-k')/N)\over \cos(2\pi(k-k')/N)}\right).
\label{fkkyz}
\end{eqnarray}
Several features of the decoherence effect are evident in the above formula.
The environment produces a tendency towards diagonalization of
the density matrix of the walker in the momentum basis (matrix elements with $k-k'=N/4$
are maximally suppressed). The decay of nondiagonal elements is exponential in time, as
already discussed in Ref. \cite{Teklemariam}. It is also clear that momentum eigenstates are not
affected by the interaction since they are eigenstates of the full
evolution [in fact, from Eqs. (\ref{fkkp}) and (\ref{superop}) follows $\tilde f_n(k,k,t)=1$].
In this sense they are perfect pointer states for this model. In what follows we will present
results concerning the evolution of several initial quantum states.

\begin{figure*}[pt]
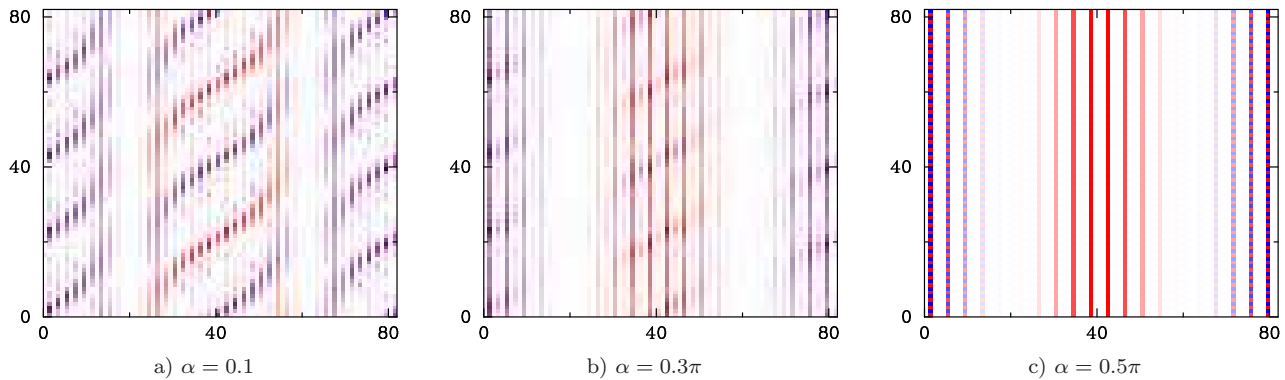

\begin{center}
\begin{minipage}[b]{0.29\textwidth}
\centering
\includegraphics[angle=270,width=1.0\textwidth]{fig5a.eps}\\
\vspace{0.2cm}
\footnotesize
a) $\alpha=0.1$
\end{minipage}%
\hspace{0.55cm}
\begin{minipage}[b]{0.29\textwidth}
\centering
\includegraphics[angle=270,width=1.0\textwidth]{fig5b.eps}\\
\vspace{0.2cm}
\footnotesize
b) $\alpha=0.3\pi$
\end{minipage}%
%\\
%\vspace{0.2cm}
\hspace{0.55cm}
\begin{minipage}[b]{0.29\textwidth}
\centering
\includegraphics[angle=270,width=1.0\textwidth]{fig5c.eps}\\
\vspace{0.2cm}
\footnotesize
c) $\alpha=0.5\pi$
\end{minipage}%
\end{center}
\caption{Discrete Wigner function $W(q,p)$ for a fixed time ($t=11$) and different values of the
strength of the coupling to the environment ($\alpha$). The initial spin is
in an unbiased state. Horizontal and vertical axis respectively correspond to position and
momentum (total dimension of the Hilbert space is $N=41$ and the initial state
is located at $n_0=20$).}
\label{wigner-local-deco}
\vspace{0.2cm}
\hrule
\end{figure*}

\begin{figure*}[hpt]
\begin{center}
\begin{minipage}[b]{0.28\textwidth}
\centering
\includegraphics[angle=270,width=1.0\textwidth]{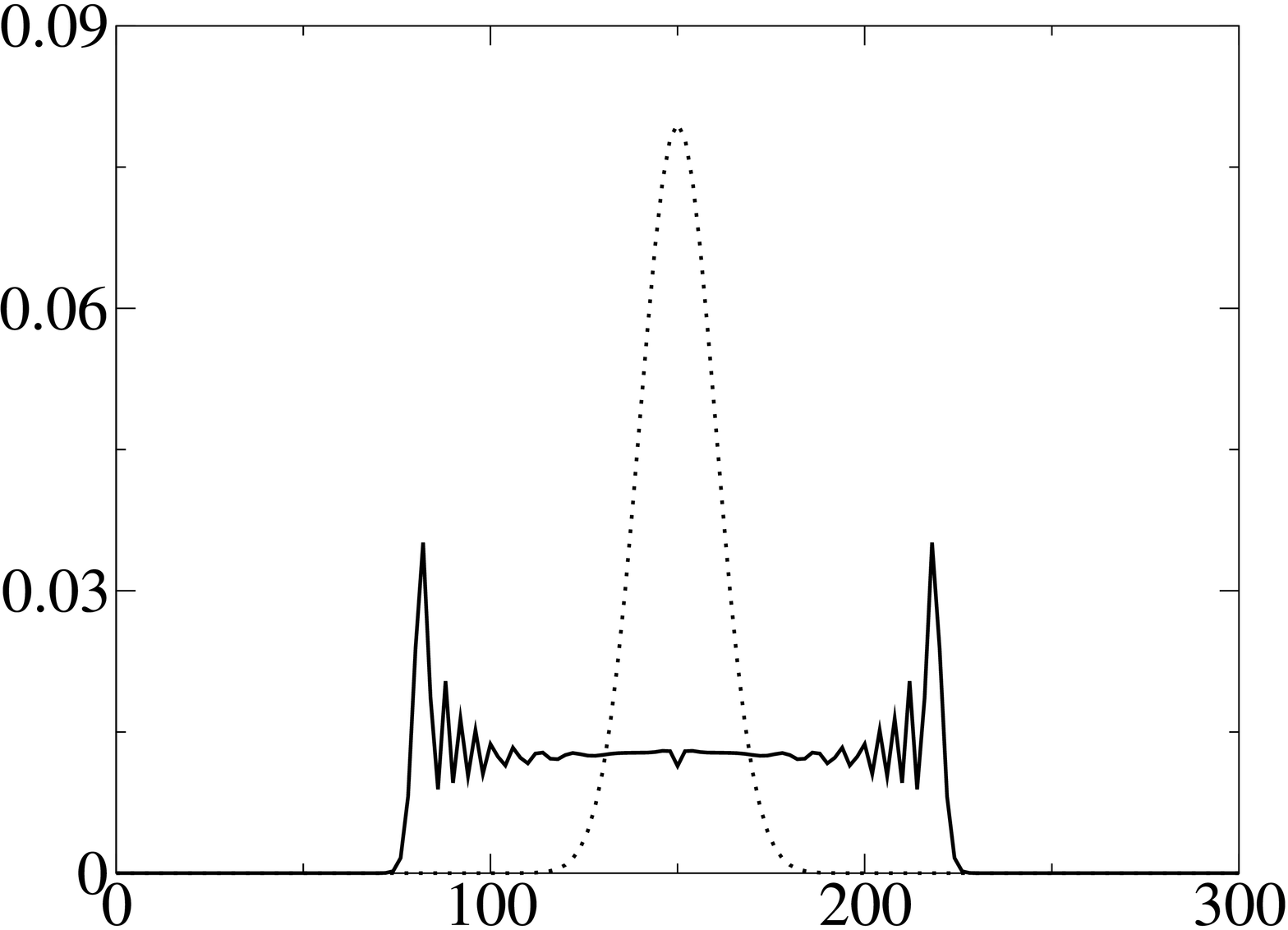}\\
\vspace{0.2cm}
\footnotesize
a) $\alpha=0.05\pi$
\end{minipage}%
\hspace{0.55cm}
\begin{minipage}[b]{0.28\textwidth}
\centering
\includegraphics[angle=270,width=1.0\textwidth]{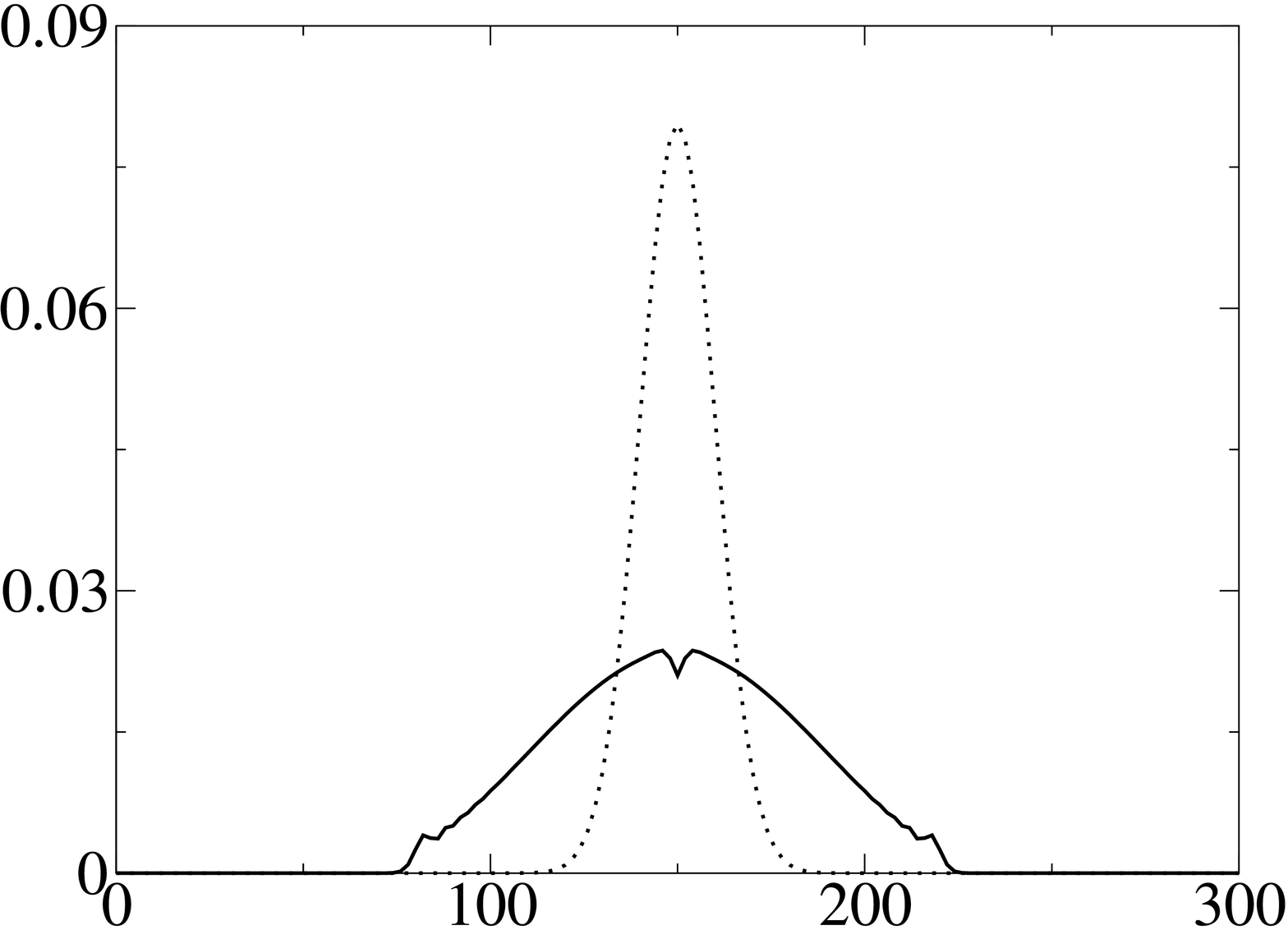}\\
\vspace{0.2cm}
\footnotesize
b) $\alpha=0.1\pi$
\end{minipage}%
\hspace{0.55cm}
\begin{minipage}[b]{0.28\textwidth}
\centering
\includegraphics[angle=270,width=1.0\textwidth]{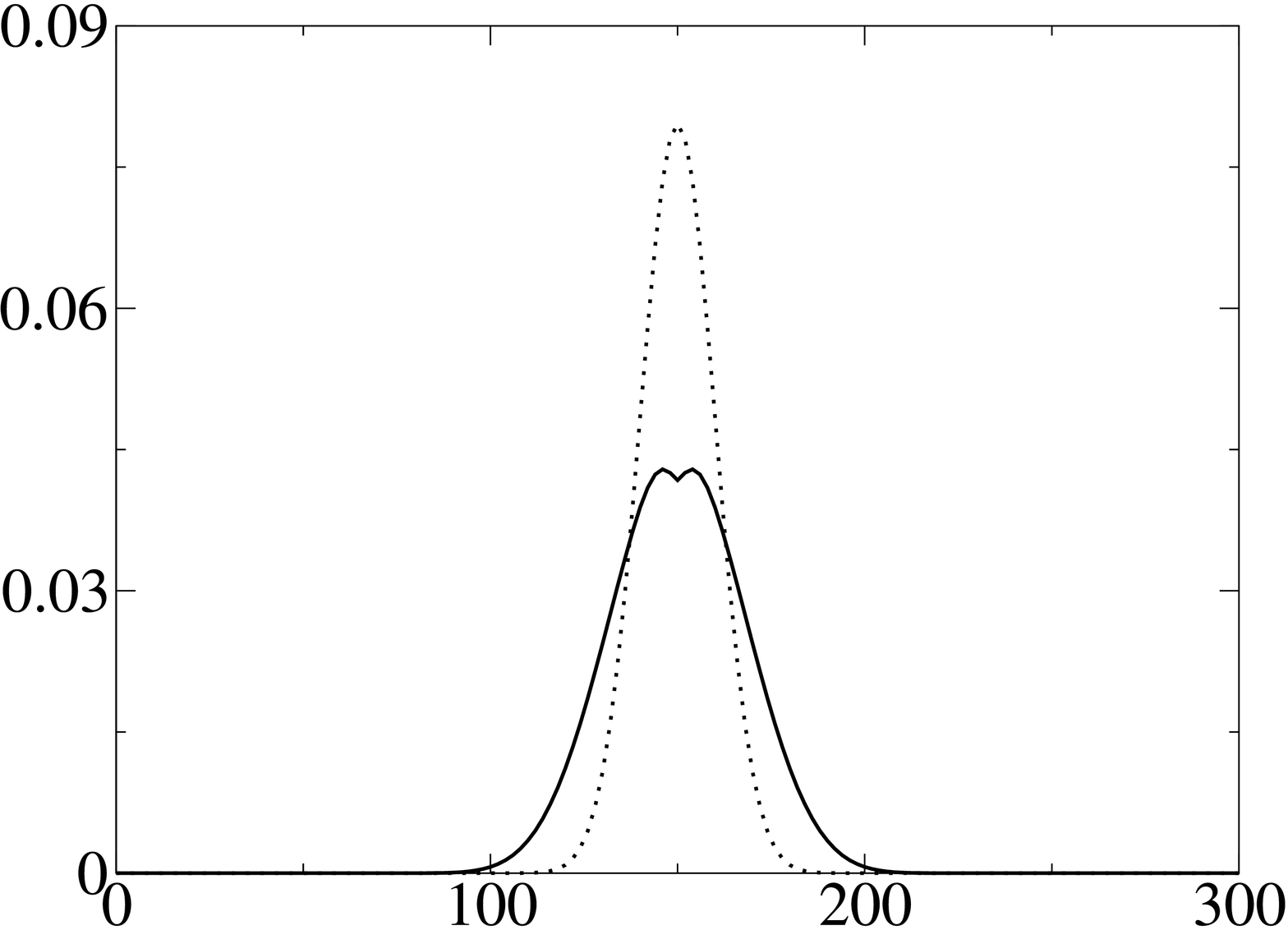}\\
\vspace{0.2cm}
\footnotesize
c) $\alpha=0.2\pi$
\end{minipage}%
\end{center}
\begin{flushleft}
%\vspace{0.2cm}
\footnotesize
\caption{Probability distribution in position at $t=100$, for different values
of the coupling to the environment (parametrized by $\alpha$). The initial state
of the quantum coin is unbiased ($N=301$, $n_0=150$, and $\hat{n}=\hat{y}$).
We only plot the function for sites such that $n+t+n_0$ adds to an even number (solid),
and also include the classical distribution (dotted).
Fig. \ref{probab-nodeco} shows the same plot without decoherence.}
\label{probab-deco}
\vspace{0.2cm}
\hrule
\end{flushleft}
\end{figure*}

\subsection{Decoherence, pointer states, and the transition from quantum to classical}

The effect of decoherence on the evolution of states which are initially localized in
position has been analyzed elsewhere \cite{KendonDec,KendonJan,BrunOct}. As shown in Fig.
\ref{wigner-local-deco}, the Wigner
function of the evolved quantum state gradually loses its oscillatory nature.
Thus, instead of a thread state the interaction with the environment gradually
produces a mixed state with a binomial distribution in the position
direction (which has an approximately Gaussian shape for large $t$) but remains constant
along the momentum direction.  It is worth noticing that for any value of $\gamma$
the resulting state has support only on position eigenstates satisfying that the sum $n_0+n+t$
is equal to an even number, as it was already pointed out for both the classical
distribution ($\gamma=0$) and the purely quantum one ($\gamma=1$) in the preceding section.
It is interesting to notice that the process of decoherence
has a rather simple interpretation when represented
in phase-space: Decoherence in phase-space is roughly equivalent to diffusion
in the position direction. This is not unexpected: In fact, in ordinary quantum Brownian motion
models a coupling to the environment through position (momentum) gives rise to a momentum
(position) diffusion term in the evolution equation for the Wigner function. The situation here
is quite similar, since the walker effectively couples with the environment through its momentum. This
is indeed the case because the environment interacts with the quantum coin which is itself
coupled with the walker through the displacement operator which is diagonal in the momentum  basis.
Therefore, the decoherence effect on the Wigner function is expected to correspond to
diffusion along the position direction.

\begin{figure*}[htp]
\begin{center}
\includegraphics[angle=270,width=0.3\textwidth]{fig7a.eps}
\hspace{0.55cm}
\includegraphics[angle=270,width=0.3\textwidth]{fig7b.eps}
\hspace{0.55cm}
\includegraphics[angle=270,width=0.3\textwidth]{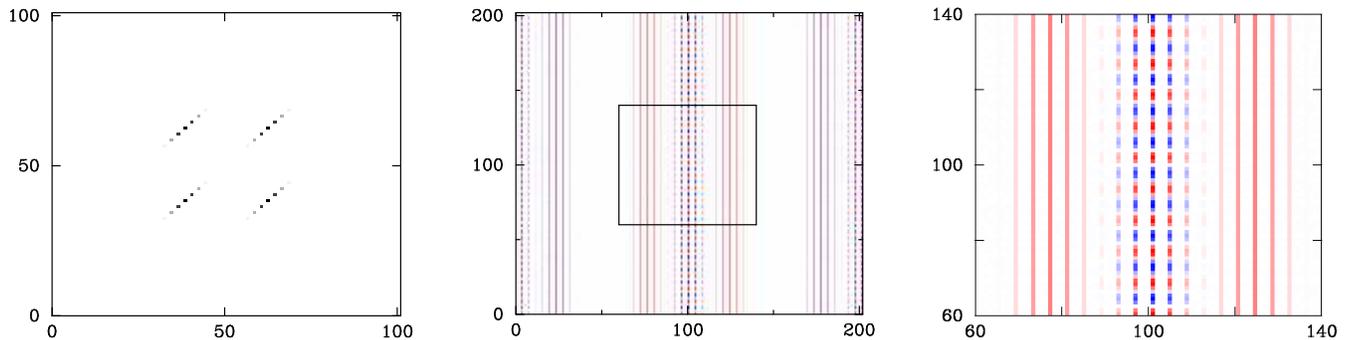}
\end{center}
\begin{flushleft}
\footnotesize
\caption{Absolute value of the density matrix in the position basis (left) and
discrete Wigner function $W(q,p)$ (center) for a state evolving from an initially
delocalized state of the quantum walker under full decoherence ($t=6$,
$\alpha=0.5\pi$, $n_1=38$, $n_2=62$, $N=101$). The presence of quantum interference
is manifested in the nondiagonal terms of the density matrix and in the oscillations
of the Wigner function. The small scale oscillations of the Wigner function
are shown in the right plot, which shows $W(q,p)$ in the phase-space region
defined by the black rectangle depicted in the center plot.}
\label{wigner-delocal-deco}
\vspace{0.2cm}
\hrule
\end{flushleft}
\end{figure*}

As noticed before, if one considers initial states where the walker is localized in a
well defined position, one can see that the probability distribution for the different
positions of the walker gradually tends to the classical one by increasing the coupling
strength $\alpha$ from $\alpha=0$ (no kicks) to $\alpha=\pi/2$. This is shown in Fig.
\ref{probab-deco}.\\

From the above analysis we could be tempted to conclude that the interaction between the
quantum coin and the environment induces the classicalization of the walker. However,
this is not the case. The process of decoherence induced in this way is not complete.
This is most clearly seen by analyzing how it is that the interaction with the environment
affects initial states of the quantum walker which are not initially localized. In
Fig. \ref{wigner-delocal-deco} we show the Wigner function of an initially delocalized
state (shown in Fig. \ref{wigner-basicas}) under full decoherence.
We can clearly see that decoherence does not erase all quantum interference effects. In fact,
as mentioned above, the interaction with the environment induces diffusion along the position
direction. Therefore, interference fringes which are aligned along the position direction are
immune to decoherence. Thus, the final state one obtains from a superposition of two position
eigenstates is not the mixture of two binomial states but a coherent superposition of them.
This peculiar behavior is easily understood by noticing that this is a simple consequence
of the fact that momentum eigenstates are pointer states: Decoherence is effective in destroying
superpositions of pointer states but highly inefficient in destroying superpositions of eigenstates
of the conjugate observable (position).

\subsection{Entropy}

\begin{figure}[ht]
\begin{center}
\includegraphics[angle=270,width=0.32\textwidth]{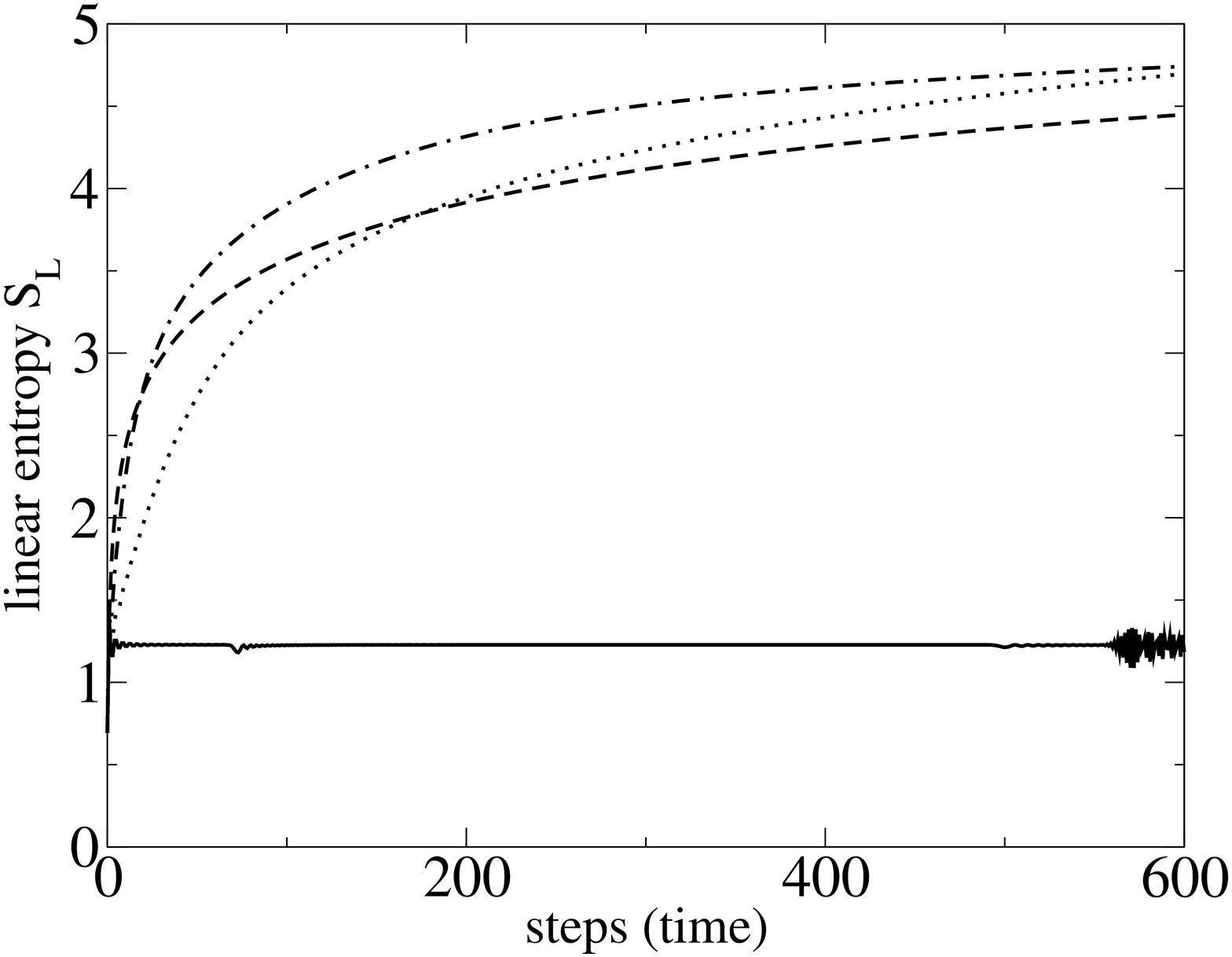}
\\
\vspace{0.25cm}
\includegraphics[angle=270,width=0.32\textwidth]{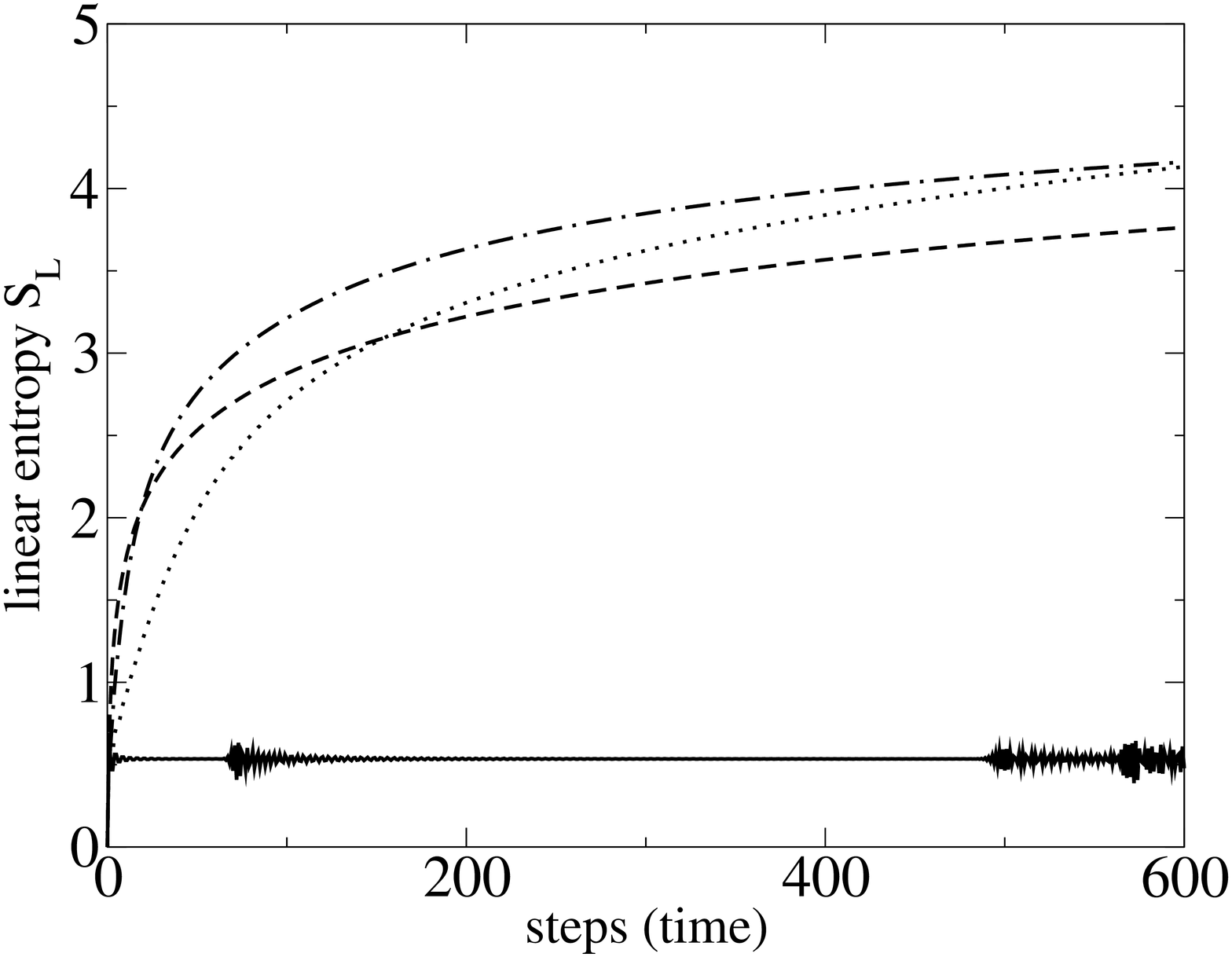}
\end{center}
\caption{Linear entropy as a function of time for various values of the system--environment
coupling strength: $\alpha = 0$ (solid), $0.1 \pi$ (dotted), $0.2\pi$ (dash dotted), $0.5\pi$ (dashed).
The top (down) plot corresponds to an initial state which is an equally weighted
mixture (superposition) of two position eigenstates. $N=401,\ n_1=150,\ n_2=250$.}
\label{entropy}
\vspace{0.2cm}
\hrule
\end{figure}

By analyzing the entropy of the reduced density matrix of the walker one can get a more quantitative
measure of the degree of decoherence achieved as a consequence of the interaction with the
environment.
For convenience we will not examine the von Neumann entropy $S_V$ but concentrate on the
linear entropy defined as $S_L=-\ln({\rm Tr}[\rho_w^2])$, which is easier to calculate. This
entropy provides a lower bound to $S_V$ \cite{entropy}. It is possible to show
that almost no entropy is produced by the decay of the coherence present in the initially delocalized
superposition state. In fact, this can be seen by comparing the entropy produced from the initially
delocalized superposition and the one originated from an initial state in which the walker is
prepared in an equally weighted mixture of two positions. These entropies can be seen in Fig.
\ref{entropy}. The initial entropy of the mixture is 1 bit [$\ln(2)$]. It is quite clear from
the curves shown in such figure that the entropy arising from the initial mixture
remains to be 1 bit higher than the one originated from the initial coherent superposition. Thus,
the quantum coherence present in the initial state is robust under the interaction with the
environment and does not decay at all.

\begin{figure}[htp]
\begin{center}
\includegraphics[angle=270,width=0.32\textwidth]{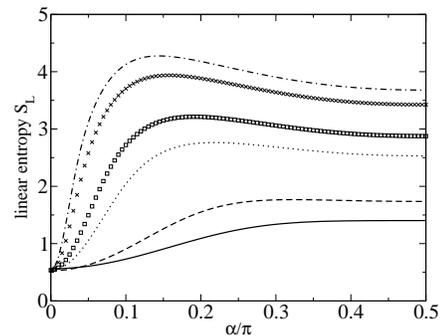}
\end{center}
\caption{Linear entropy as a function of the coupling strength $\alpha$ for various
values of time: $t=5$ (solid), $10$ (dashed), $50$ (dotted), $100$ ($\square$),
$300$ ($\times$), $500$ (dash dotted).
The initial state of the walker is well localized and the initial
state of the quantum coin is unbiased. $N=401$ and all the curves are below the
saturation regime [$\ln(N)\approx 5.994$]. It is evident that, after some time, the
maximum value of the entropy is not achieved by the maximum value of the coupling
strength.}
\label{entropy-rate}
\vspace{0.2cm}
\hrule
\end{figure}

Fig. \ref{entropy} shows another interesting feature: One would
naively expect a monotonic dependence of the entropy with the coupling to the
environment (which is parametrized by $\alpha$). However this is not the case
since the curves in Fig. \ref{entropy} intersect. This peculiar effect is
made more evident in Fig. \ref{entropy-rate} where we study the entropy at
a fixed time as a function of the coupling strength. In this figure a clear
indication of a nontrivial behavior is seen: For early times the entropy grows slowly
with $\alpha$ and exhibits a flat plateau for large values of $\alpha$. However,
as time progresses a peak develops: The largest value of entropy at a given time
is not achieved by the largest coupling. To the contrary, the largest entropy is
attained by an intermediate coupling $\alpha_c$, whose value decreases with time.

The fact that for a given time the maximal entropy is not achieved by the maximal
coupling to the environment is counterintuitive. As entropy is a measure of
the spread of a distribution, this strange behavior can be rephrased as a
manifestation of the counterintuitive fact that the decohered state (which is
approximately diagonal in position basis) has a probability distribution that is
more spread for $\alpha=\alpha_c$ than for $\alpha>\alpha_c$.
A possible explanation for this peculiar behavior is the following: For high
values of the coupling to the environment the state rapidly becomes classical
and the spread in position grows diffusively, as in the classical
random walk. When the coupling to the environment is not strong, our result
seems to indicate that the state of the walker remains ``quantum'' for a
longer time during which it spreads at a rate faster than classical. When
this quantum state finally decoheres it may end up having a larger entropy
than the one attained for high coupling simply because it is spread over
a wider range of positions.
We speculate that there could be a relation between this peculiar feature
and the properties that make some degree of decoherence useful for quantum walks as
discussed by Kendon and Tregenna in \cite{KendonDec,KendonJan}. The value of the $\alpha_c$
introduced above depends on both $N$ and $t$ and could be related to
the position of the minima reported in Refs. \cite{KendonDec,KendonJan}.
For example, in a cycle regime ($t >> N$) $\alpha_c$ diminishes with increasing $N$
as it is also the case for the position of the minima of the so--called
\textit{quantum mixing time} \cite{Kempe,KendonDec,KendonJan}.

\section{Conclusions}

The use of phase-space representation enables us to develop some intuition about the nature
of the decoherence process in the kind of quantum walk analyzed in this paper. By coupling
the quantum coin to an environment we obtain a decoherence model which is roughly equivalent
to position diffusion. As we mentioned above, this is a natural result whose origin can
be traced back to the way in which the system effectively couples to the environment (via
the momentum operator). The relation between decoherence and position diffusion can also
be established by analyzing in more detail the structure of the superoperator $O_n$
[given in Eq. (\ref{superop})]. Let us consider the form of the superoperator after
$t$ iterations. If we use Eq. (\ref{superop}) we can easily see that, as each iteration
doubles the number of terms, we will have an expression with $2^t$ terms each one of which
has Pauli operators applied at different times. To obtain the
function $\tilde f_n(k,k',t)$ one should compute the trace over the quantum coin. In each of
the $2^t$ terms we can move the Pauli operator $\sigma_n$ towards the outside of the
expression and cancel them due to the cyclic property of the trace. For the case $\hat n=\hat y$
it is easy to show that the only remaining effect of the Pauli operators (that in this case
anti--commute with the Hadamard operator) is to reverse the direction of the rotation in $M_k$
defined in Eq. (\ref{Mofk}). The final expression can be shown to be
\begin{eqnarray}
{\rm Tr}(O_y^t(\rho))&=&\sum_{\alpha_t=0,1}\ldots\sum_{\alpha_1=0,1} {1\over 2^t}
\left(1+\gamma\right)^{t-\bar\alpha_t}
\left(1-\gamma\right)^{\bar\alpha_t}\nonumber\\
&\times&{\rm Tr}\left(M_{k_t}\ldots M_{k_1}\rho M_{k'_1}^\dagger\ldots M_{k'_t}^\dagger\right),
\label{super2}
\end{eqnarray}
where $\bar\alpha_j=\alpha_1+\ldots +\alpha_j$ and $k_j=(-1)^{\bar\alpha_{j-1}}k$ (we use
the convention $\bar\alpha_0 \equiv 0$).
Therefore, the superoperator is the sum of $2^t$ terms each one of which contains
a contribution that is identical to that of a quantum walk where the direction of the
walker is chosen at random after the first step. Each of the $2^t$ terms is labeled by
a $t$--bit string $(\alpha_1,\ldots,\alpha_n)$ and corresponds to a quantum walk where
the direction of the $j$th step ($j\ge 2$) is reversed if and only if $\bar\alpha_{j-1}$ is odd.
In the limit of total decoherence each of these terms has equal weight. Therefore the final
state is simply the average over an ensemble where each member corresponds to each
of the $2^{t-1}$ possible choices of two directions (forward or backward) for the $t-1$ steps
(notice that the direction of the first step is not affected by the decoherence model
we chose). For this type of decoherence it is clear that the quantum walk becomes a
random walk. The relation between decoherence and position diffusion is quite evident in this
way. It is worth pointing out that similar models of decoherence
were considered in Ref. \cite{Bianucci} in a different context.

Other decoherence models have been analyzed for quantum
walks \cite{KendonDec,KendonJan,Dur}, where the effective coupling to the environment is through the position
observable. In such case, we expect decoherence to correspond to diffusion along the momentum
direction. Combining the two types of decoherence (i.e., considering
coupling to the environment via the quantum coin and the position of the quantum walker) the
initial state corresponding to a superposition of two positions would finally decay into
a mixture of two binomial states (see Ref. \cite{Bianucci} for similar results obtained when
studying decoherence models with a natural phase-space representation in a finite quantum system
evolving under various quantum maps).

The above conclusions are generic for any model in which decoherence is due to the coupling
of the quantum coin to an environment. An interesting class of models, based on the use of
quantum multi--Baker maps \cite{multibaker}, has been studied. In such models one replaces the
quantum coin with a quantum system with a higher-dimensional Hilbert space. The total space
of states is then $\mathcal{H}=\mathcal{H}_N \otimes \mathcal{H}_M$. Here $M$ is the dimensionality
of the system which plays the role of the quantum coin and is considered to be an even number
($M=2m$, so that we can always consider $\mathcal{H}_M=\mathcal{H}_2 \otimes \mathcal{H}_m$).
The dynamics for a quantum multi--Baker map is defined in terms of the unitary operator [that
replaces Eq. (\ref{onestep})]:
\begin{equation}
\mathcal{U}_{multi-Baker} = U^{\sigma_z}\  B_M,
\label{mbaker}
\end{equation}
where $B_M$ is the unitary operator defining the so--called ``quantum Baker map'' (see
Refs. \cite{Saraceno,Bianucci}) and $\sigma_z$ is a Pauli operator acting on the Hilbert space
$\mathcal{H}_2$ (the most significant qubit of the internal space
$\mathcal{H}_M=\mathcal{H}_2 \otimes \mathcal{H}_m$). The properties of the operator $B_M$
have been widely studied in the literature \cite{Saraceno}: The map faithfully represents a
classically chaotic system (in the large $M$ limit). From the point of view of the quantum
walker the situation is quite similar to the one we studied in this paper. One can describe
a quantum multi--Baker map as an ordinary quantum walk where the quantum coin (whose
Hilbert space is $\mathcal{H}_2$) interacts with an environment (whose Hilbert space
is $\mathcal{H}_m$). The interaction is modeled by the quantum Baker map acting on the
total internal Hilbert space $\mathcal{H}_M$, which also replaces the usual Hadamard
step in Eq. (\ref{onestep}). As the quantum Baker map is chaotic, the state of the quantum
coin will be roughly randomized after each iteration. Thus, the effect should be similar
to the one we described here (where the quantum coin is subject to a noisy evolution).
However, after a large number of iterations (of the order of $M$) all the possible orthogonal
directions available in the internal space of the quantum coin would have been explored. One
should therefore expect that this model will stop being effective in producing decoherence
after such time. Recent studies of quantum multi--Baker systems agree with these expectations
(see Ref. \cite{multibaker} where a transition from diffusive to ballistic behavior after a
time of the order of $M$ has been analyzed).
In any case, based on the results of our work we believe that in quantum multi--Baker systems
the relative stability of initially delocalized superpositions will also be observable.

We thank Marcos Saraceno and Augusto Roncaglia for useful discussions and assistance
during several stages of this work. C.C.L. was partially supported by funds
from Fundaci\'on Antorchas, Anpcyt and Ubacyt.\\

\appendix

\section{Density matrix under decoherence}

A matrix representation for the superoperator $O_n$ can be obtained for
$\hat{n}=\hat{x},\hat{y},\hat{z}$. We will write this superoperator in the
basis formed by the identity and the Pauli matrices (we use the standard ordering
of the basis as $\{I,\ \sigma_x,\ \sigma_y,\ \sigma_z\}$). In such basis the
matrix of $O_n$ is
\begin{eqnarray*}
O_y =
\left (
\begin{array}{cccc}
\cos(-) & -i \sin(-) & 0 & 0 \\
0 & 0 & \gamma \sin(+) & \gamma \cos(+) \\
0 & 0 & -\cos(+) & \sin(+) \\
-\gamma i \sin(-) & \gamma \cos(-) & 0 & 0 \\
\end{array}
\right ),
%\label{defOy}
\end{eqnarray*}
where [$\pm)=(2\pi(k \pm k')/N$].

Similar results are obtained for $\hat{n}=\hat{x}$, $\hat{z}$:
\begin{eqnarray*}
O_x =
\left (
\begin{array}{cccc}
\cos(-) & -i \sin(-) & 0 & 0 \\
0 & 0 & \sin(+) & \cos(+) \\
0 & 0 & -\gamma \cos(+) & \gamma \sin(+) \\
-\gamma i \sin(-) & \gamma \cos(-) & 0 & 0 \\
\end{array}
\right ),
\end{eqnarray*}

\begin{eqnarray*}
O_z =
\left (
\begin{array}{cccc}
\cos(-) & -i \sin(-) & 0 & 0 \\
0 & 0 & \gamma \sin(+) & \gamma \cos(+) \\
0 & 0 & -\gamma \cos(+) & \gamma \sin(+) \\
-i \sin(-) & \cos(-) & 0 & 0 \\
\end{array}
\right ).
\end{eqnarray*}
Notice they all converge to the same matrix when $\gamma=1$ (no decoherence).\\

To compute Eq. (\ref{fkkplanteodeco}) we need to diagonalize $O_n$. Although the
matrix representation of the superoperator is rather sparse, the eigenvalues and
eigenvectors are quite cumbersome for an arbitrary value of $\gamma$ (including
$\gamma=1$, no decoherence), so it was more convenient to use numerical techniques.
However, for the special case of complete decoherence ($\gamma=0$)
it is possible to obtain a simple formula and the final result
for $\tilde f_n(k,k',t)$. The result for $\hat n=\hat y, \hat z$ is given in
Eq. (\ref{fkkyz}).

\end{document}